\documentclass[epj,nopacs]{svjour}
\usepackage{amssymb}
\usepackage{graphicx}

\def\Vec#1{{\bf #1}}

\def\D{{\mathrm d}}

\providecommand{\Journal}[4] {#1 {\bf #2} (#4) #3}
\providecommand{\ZPC}{Z. Phys. C } %

\begin{document}

\title{The $\cos 2 \phi$ asymmetry of Drell--Yan and 
$J/\psi$ production in unpolarized
$p\bar{p}$ scattering}

\author{Vincenzo Barone\inst{1}\and Zhun Lu\inst{2,3}\and 
Bo-Qiang Ma\inst{2}}

\institute{Di.S.T.A., Universit\`a del Piemonte Orientale
``A. Avogadro'', \\ and INFN, Gruppo Collegato di Alessandria,
15100 Alessandria, Italy \and 
School of Physics, Peking University, Beijing 100871, China \and
Departamento de F\'\i sica, Universidad T\'ecnica Federico
Santa Mar\'\i a, Casilla 110-V, Valpara\'\i so, Chile}

\mail{V.~Barone, barone@to.infn.it; B.-Q.~Ma, mabq@phy.pku.edu.cn}

\abstract{We investigate the $\cos 2 \phi$ azimuthal asymmetry in 
Drell--Yan and $J/\psi$
production from unpolarized $p\bar{p}$ scattering at GSI-HESR
energies. The contribution to this asymmetry arising from the
leading-twist Boer-Mulders function $h_1^{\perp}(x, \Vec k_T^2)$,
which describes a correlation between the transverse momentum and
the transverse spin of quarks in an unpolarized hadron, is
explicitly evaluated, and predictions for the GSI-HESR kinematic
regime are presented. We show that the $\cos 2 \phi$ asymmetry is
quite sizable both on the $J/\psi$ peak and in the Drell-Yan
continuum region.
Therefore these processes may 
offer an experimentally viable access to the
Boer-Mulders function in the early unpolarized stage of GSI
experiments.}

\maketitle


\section{Introduction}

The spin structure of hadrons is under intense investigation, both
theoretically and experimentally. In hadron--hadron scattering one
would naively speculate that the polarization of at least one of the
incoming hadrons, the beam or the target, or of both, is needed in
order to investigate the spin properties of hadrons. However this is
not the case if one takes into account the intrinsic momentum of
partons. It is known in fact that there is a new kind of
leading-twist transverse-momentum dependent distribution function,
the so called Boer--Mulders function $h_1^{\perp}(x, \Vec k_T^2)$
\cite{bm,boer}, which describes the transversity of quarks  inside
an unpolarized hadron. This new function, which is chirally odd,
manifests itself through the coupling to another chiral-odd
quantity, such as the transversity distribution $h_1$, the Collins
function $H_1^{\perp}$, or another Boer-Mulders function \cite{bdr}.
Therefore the spin structure of hadrons can be also studied in
physical processes without beam and/or target polarization. It has
been argued by Boer~\cite{boer} that $h_1^{\perp}(x, \Vec k_T^2)$
can account for the large $\cos 2\phi$ azimuthal asymmetry observed
in the unpolarized pion-nucleon Drell-Yan
process~\cite{na10,conway}. The origin of $h_1^{\perp}(x, \Vec
k_T^2)$ has been addressed in Refs.~\cite{collins02,bjy02}. The
first calculation of $h_1^{\perp}(x, \Vec k_T^2)$, in a
quark--scalar diquark model, was reported by Goldstein and Gamberg
\cite{gg02}. Soon after this work, Boer, Brodsky, and
Hwang~\cite{bbh03}, using a similar model, showed that the analyzing
power $\nu$ of the $\cos 2\phi$ asymmetry could be as high as 30\%
in the unpolarized $p\bar{p}$ Drell-Yan process, and smaller in $pp$
scattering. The Boer--Mulders function of the pion was
calculated~\cite{lm04,lm05} in a quark--spectator-antiquark model
and shown to reproduce, in combination with the corresponding
distribution of the proton~\cite{bsy04}, the $\cos 2\phi$ asymmetry
in the $\pi^- N$ Drell-Yan process measured by NA10
Collaboration~\cite{na10}. The Boer-Mulders function, combined with
the Collins function, can also produce the $\cos 2\phi$ asymmetries
of semi-inclusive pion leptoproduction from unpolarized
nucleons~\cite{gg03,BLM05}.

Recently, the $\cos 2\phi$ azimuthal asymmetry of the $p\bar{p}$
Drell-Yan process has received much attention, as there have been
proposals to study spin phenomena in polarized and unpolarized
$p\bar{p}$ scattering at the High-Energy Storage Ring (HESR) of
GSI \cite{PAX,PANDA}. There have been numerical
simulations~\cite{MC-DY} as well as model
calculations~\cite{bbh03,GG05} of the $\cos 2\phi$ asymmetry in
the unpolarized $p\bar{p}$ Drell-Yan process. The purpose of this
paper is to investigate the $\cos 2\phi$ asymmetry of 
Drell--Yan and $J/\psi$ production in 
unpolarized $p \bar p$ scattering, using 
a model of $h_1^{\perp}(x, \Vec k_T^2)$ \cite{lm05,bsy04,BLM05}
that has been 
adjusted to fit the NA10 $\pi^- N$ Drell-Yan data~\cite{na10} and
shown to be 
in agreement with the  ZEUS~\cite{zeus} and EMC~\cite{EMC}
semi-inclusive DIS data. As we will see,
 in the GSI kinematic domain the $\cos 2\phi$  asymmetry turns out to be
rather large. In particular, 
the advantage of studying the $J/\psi$ production is
that, while its $\cos 2 \phi$ asymmetry is similar in size to the
corresponding asymmetry of the Drell--Yan continuum production,
the counting rate on the $J/\psi$ peak is two orders of magnitude
higher compared to the region above this peak.

\section{The $\cos 2 \phi$ asymmetry in Drell-Yan and 
$J/\psi$ production}

The Drell-Yan process represents an ideal window on the hadron
structure, since it probes only parton densities. We are
interested in the transverse momentum distribution of the lepton
pairs, hence we have to consider the intrinsic transverse momenta
of partons inside the hadrons. The non-collinear factorization
theorem for Drell-Yan process has been recently proven by Ji, Ma
and Yuan \cite{Ji:2004xq} for $Q_{T} \ll Q$.

The angular differential cross section for the unpolarized
Drell-Yan process is usually parametrized as
\begin{eqnarray}
& & \frac{1}{\sigma^{\rm DY}}\frac{d\sigma^{\rm DY}}{d\Omega}=
\frac{3}{4\pi}\frac{1}{\lambda+3} \left (1+\lambda\cos^2\theta
\right. \nonumber \\
& & \hspace{0.5cm} \left. +\mu
\sin2\theta\cos\phi+\frac{\nu}{2}\sin^2\theta\cos2\phi\right )\,
.\label{angular}
\end{eqnarray}
where $\theta$ and $\phi$ are, respectively, the polar angle and
the azimuthal angle of dileptons in the Collins--Soper 
frame \cite{cs77}. At leading order, the $\phi$-independent term of the
unpolarized Drell-Yan cross section for proton-antiproton
collisions is
\begin{eqnarray}
& & \frac{d\sigma_{p\bar p}^{\rm DY}}{d\Omega \, dx_1 \, d x_2 \,
d^2\Vec q_T} = \frac{\alpha^2_{\rm em}}{12 \, M^2} (1 + \cos^2
\theta) 
\nonumber \\
& & \hspace{0.5cm} \times \sum_{a}e_a^2
\int d^2\Vec k_T \, d^2\Vec p_T \, 
\delta^2(\Vec k_T+\Vec p_T-\Vec q_T) 
\nonumber \\
& & \hspace{0.5cm} \times f_1^a(x_1,\Vec k_T^2)
f_1^a(x_2,\Vec p_T^2),\label{cs}
\end{eqnarray}
where $M^2$ is the invariant mass squared of the lepton pair and
$\Vec q_T$ is the transverse momentum of the virtual photon in the
frame where the two colliding hadrons are collinear. In
Eq.~(\ref{cs}) we denoted by $f_1 (x, \Vec k_T^2)$ the
unintegrated quark number density in the proton, and we used the
fact that the antiquark distributions in the antiproton are equal
to the quark distributions in the proton.

We will consider the $\nu$ term in (\ref{angular}). It is known 
that gluon radiation processes give rise to a non-zero 
$\cos 2 \phi$ asymmetry, which
in case of $q \bar q$ annihilation dominance is given by 
$\nu = Q_T^2/(M^2 + 3 Q_T^2/2)$ \cite{collins79}. 
Various quantitative analyses \cite{chiappetta86,brandenburg93,na10} 
show that perturbative corrections are unable to reproduce both the 
magnitude and the $Q_T$-dependence of $\nu$ 
as observed by NA10 in the region $M \sim 4-8$ GeV,  but 
for lower $M$ values the perturbative asymmetry might be 
relevant. 
In the present paper we 
will focus on the 
Boer--Mulders contribution to the $\cos 2 \phi$ 
asymmetry, which has been shown to explain the NA10 results \cite{boer}
and represents a large 
effect in the moderate--$Q_T$ 
region we are interested in.\footnote{Concerning the 
perturbative contributions to azimuthal
asymmetries, a forthcoming analysis 
of $\cos 2 \phi$ distributions in semiinclusive DIS 
shows that at moderate $Q^2$ and low $P_T$ the Boer--Mulders effect 
largely dominates over the perturbative contribution
(at large $Q^2$ the situation is reversed) \cite{barone07}.}

The expression for the contribution of the Boer-Mulders
distribution to the unpolarized cross-section is~\cite{boer}
\begin{eqnarray}
& & \left. \frac{d\sigma_{p\bar{p}}^{\rm DY}}{d\Omega \, dx_1 \, d x_2
\, d^2\Vec q_T} \right \vert_{\cos 2 \phi}=\frac{\alpha^2_{\rm
em}}{12 \, M^2} \sin^2 \theta 
\nonumber \\
& & \hspace{0.5cm} \times \sum_a e_a^2\int d^2\Vec k_T
d^2\Vec p_T \, 
 \delta^2(\Vec k_T+\Vec p_T-\Vec q_T)
\nonumber \\
& & \hspace{0.5cm} \times 
\frac{(2\hat{\Vec
h}\cdot \Vec k_T\hat{\Vec h}\cdot \Vec p_T
-\Vec k_T\cdot \Vec p_T)}{m_N^2}\nonumber\\
&&\hspace{0.5cm} 
\times \, h_1^{\perp a}(x_1,\Vec k_T^2) h_1^{\perp a}(x_2,\Vec
p_T^2)\cos2\phi ,\label{bm}
\end{eqnarray}
where $\hat{\Vec h} \equiv \Vec q_T/Q_T$, with $Q_T \equiv \vert
\Vec q_T \vert$. This term, with its peculiar angular dependence,
gives a parton model explanation for the $\cos 2 \phi$ asymmetry
observed in the unpolarized $\pi N$ Drell-Yan process \cite{na10}.

From Eqs.~(\ref{cs}) and (\ref{bm}) we get the following
expression of the coefficient $\nu$ in (\ref{angular}) (with
$\lambda=1$, $\mu=0$) :
\begin{equation}
\nu=\frac{2\sum_{a}e_a^2\, \mathcal{H}[h_1^{\perp a}, h_1^{\perp
a}]}{ \sum_{a}e_a^2\, \mathcal{F}[f_1^a, f_1^{a}]}, \label{nu}
\end{equation}
where we used the notations:
\begin{eqnarray}
 \mathcal{F}[f_1^a, f_1^{a}] &=& \frac{1}{M^2} \, \int d^2\Vec k_T 
\, d^2\Vec
p_T\delta^2(\Vec k_T+\Vec p_T-\Vec q_T)
\nonumber \\
& &  
\times f_1^a(x_1,\Vec k_T^2) f_1^{a}(x_2,\Vec p_T^2)\nonumber\\
& =& \frac{1}{M^2} \, 
\int \D k_T \, k_T \, \int_0^{2 \pi} \D \chi 
\nonumber \\
& & \times 
 f_1^{a}(x_1,
\Vec k_T^2) \, f_1^{a}(x_2, \vert \Vec q_T -
\Vec k_T \vert^2)\,,\label{fconv}
\end{eqnarray}
\begin{eqnarray}
 \mathcal{H}[h_1^{\perp a},h_1^{\perp a}] &=& \frac{1}{M^2} 
\, \int d^2\Vec k_T
\, d^2\Vec p_T\delta^2(\Vec k_T+\Vec p_T-\Vec q_T)\nonumber\\
&& 
\times\frac{(2\hat{\Vec h}\cdot \Vec k_T\hat{\Vec h}\cdot \Vec
p_T -\Vec k_T\cdot \Vec p_T)}{M^2} 
\nonumber \\
& &  \times h_1^{\perp a}(x_1,\Vec k_T^2)
h_1^{\perp a}(x_2,\Vec p_T^2)
\nonumber \\
& =& \frac{1}{M^2} \, 
\int \D k_T \, k_T \, \int_0^{2 \pi} \D \chi 
\nonumber \\
& &  \times 
 \frac{\Vec
k_T^2+ Q_T\, k_T \, \cos \chi - 2 \, \Vec k_T^2 \, \cos^2 \chi
}{m_N^2}
\nonumber \\
& &  
\times \, h_1^{\perp a}(x_1, \Vec k_T^2) \,h_1^{\perp a}(x_2,
\vert \Vec q_T - \Vec k_T \vert^2)\, ,\label{hconv}
\end{eqnarray}
and $\chi$ is the angle between $\Vec q_T$ and $\Vec k_T$. The
asymmetry coefficient $\nu$ given in Eq.~(\ref{nu}) depends on the
kinematic variables $x_1$, $x_2$, $M$ and $Q_T$. The GSI-HESR
kinematics probes the large $x$ region, where the valence quarks
and in particular the $u$ sector dominate, so that Eq.~(\ref{nu})
can be simplified to:
\begin{equation}
\nu=\frac{2\mathcal{H}[h_1^{\perp u}, h_1^{\perp u}]}{
\mathcal{F}[f_1^u, f_1^{u}]}.\label{nu2}
\end{equation}

Another promising process to study transverse spin physics at
GSI-HESR is $J/\psi$ production, which has been proposed
\cite{abdn} as a method to access transversity by measuring double
spin asymmetry. The $J/\psi$ production events are identified by a
peak in the dilepton invariant mass spectrum at
$M=m_{J/\psi}=3.097$~GeV, so one can choose $M^2 \simeq
9$~GeV$^2$. At the GSI-HESR energy scale ($ s = 30-45$ GeV$^2$ in
the fixed target mode, or $s = 100-200$ GeV$^2$ in the collider
mode) $J/\psi$ production is dominated by $q \bar q$ annihilation
\cite{jpsi}. Since the $J/\psi$ is a vector particle and the $q
\bar q J/\psi$ coupling has the same helicity structure as the $q
\bar q \gamma^*$ coupling, one can get the $J/\psi$ production
cross section by replacing the
 quark electric charges by the
 $J/\psi$ vector couplings to $q \bar q$ and to $\ell^+ \ell^-$,
\begin{equation}
16 \pi^2 \alpha_{\rm em}^2 e_a^2 \rightarrow (g_a^V)^2
(g_{\ell}^V)^2 \,,
\end{equation}
and the virtual photon propagator by a Breit--Wigner function,
\begin{equation}
\frac{1}{M^4} \rightarrow \frac{1}{(M^2 - m_{J/\psi}^2)^2 +
m_{J/\psi}^2 \Gamma_{J/\psi}^2}\,,
\end{equation}
where $\Gamma_{J/\psi}$ is the $J/\psi$ width. This model
successfully accounts for the SPS $J/\psi$ production data at
moderate $s$ \cite{sps} .

Therefore the unpolarized $\cos 2\phi$ asymmetry in the $J/\psi$
resonance region reads
\begin{equation}
\nu=\frac{2\sum_{a}(g_a^V)^2\mathcal{H}[h_1^{\perp a}, h_1^{\perp
a}]}{ \sum_{a}(g_a^V)^2\mathcal{F}[f_1^a, f_1^{a}]},
\end{equation}
with only $e_a^2$ replaced by $(g_a^V)^2$ in Eq.(\ref{nu}). Since
the $u$ quark dominates, this asymmetry reduces to
Eq.~(\ref{nu2}), which is the same as for continuum Drell-Yan,
with $M^2\simeq 9$~GeV$^2$. Integrating over $x_1$ and $x_2$ we
get
\begin{equation}
\nu(Q_T,M)=\frac{2\int d x_1 \int d x_2 \mathcal{H}[h_1^{\perp
u},h_1^{\perp u}]\, \delta(1- \tau)}{\int d x_1 \int d x_2 \,
\mathcal{F}[f_1^u, f_1^{u}]\, \delta(1- \tau)},\label{nu3}
\end{equation}
where $\tau=M^2/ x_1 x_2 s$. Notice that the relation $1- \tau=0$
is valid for $Q_T ^2\ll M^2$. The rapidity is defined as
$y=\frac{1}{2}\, \ln (\frac{x_1}{x_2})$ and Feynman's variable is
 $x_F=x_1 - x_2$.

\section{Calculation of the $\cos 2 \phi$ asymmetry}

To evaluate the  $\cos 2 \phi$ asymmetry given in (\ref{nu2}) one
needs to know the form of the $k_T$-dependent distributions
appearing in the transverse momentum convolution. Useful
information on the Boer-Mulders function $h_1^{\perp}(x, \Vec
k_T^2)$ can be obtained from the study of the $\cos 2 \phi$
azimuthal asymmetry in the unpolarized $\pi N$  Drell-Yan
processes, which has been measured by the NA10 Collaboration
\cite{na10} and the E615 Collaboration \cite{conway}. In
\cite{lm04,lm05} this asymmetry was estimated by computing the
$h_1^{\perp}$ distribution of the pion and of the nucleon in a
quark spectator model \cite{bsy04,jmr97} and was compared with
NA10 data. To compute the $\cos 2 \phi$ azimuthal asymmetry in the
unpolarized $p\bar{p}$ collision we adopt the same distributions
$h_1^{\perp} (x, \Vec k_T^2)$ and $f_1 (x, \Vec k_T^2)$ used in
\cite{lm05}. We assume that the observables are dominated by $u$
quarks. The set of the transverse-momentum dependent distribution
functions is (for simplicity, we consider a spectator scalar
diquark \cite{lm05,bsy04})
\begin{eqnarray}
 f_{1}^{u}(x,\Vec k_T^2) &=& N \, (1 - x)^3 \, \frac{(x m_N +
m_q)^2 + \Vec k_T^2}{(L^2 + \Vec k_T^2)^4},
\label{f1} \\
 h_{1}^{\perp u}(x, \Vec k_T^2) &=& \frac{4 \alpha_s}{3} \,
N\, (1 - x)^{3} \, \frac{m_N \, (x m_N + m_q)}{[L^2 \, (L^2 + \Vec
k_T^2)^3]}, \label{h1perp}
\end{eqnarray}
where $N$ is a normalization constant, $m_q$ is the constituent
quark mass, and
\begin{equation}
L^2=(1 - x) \, \Lambda^2 + x \, m_d^2  - x \, (1 - x) \, m_N^2\,.
\end{equation}
Here $\Lambda$ is a cutoff appearing in the nucleon-quark-diquark
vertex and $m_d$ is the mass of the scalar diquark. As it is
typical of all model calculations of quark distribution functions,
we expect that Eqs.(\ref{f1}) and (\ref{h1perp})
 should be valid at low $Q^2$ values, of order of 1 GeV$^2$.
The average transverse momentum of quarks inside the nucleons, as
computed from (\ref{f1}), turns out to be $\langle k_T^2
\rangle^{1/2} \simeq 0.54$ GeV. For the parameters in
Eqs.(\ref{f1}) and \ref{h1perp}), we choose the values $m_d=0.8$
GeV, $m_q=0.3$ GeV, $\Lambda=0.6$ GeV, $\alpha_s=0.3$, which are
the same as in \cite{lm05,BLM05}.

Compared with previous calculations of Drell-Yan
asymmetries~\cite{bbh03,GG05}, our model differs for the explicit
form of the proton-quark-diquark effective coupling and for the
values of the parameters. We adopt a dipole form
factor~\cite{bsy04} for the effective coupling, whereas 
in other computations this
coupling is taken as a constant~\cite{bbh03}, 
or as a Gaussian form factor~\cite{GG05}. This leads to different
predictions on the magnitude and the $Q_T$-dependence of the $\cos
2\phi$ asymmetry, as we will see below. 
Our model proved to be capable to reproduce the
observed $\cos 2\phi$ asymmetry of the NA10 $\pi^- N$ Drell-Yan
data with the correct $Q_T$-dependence~\cite{lm05}.

The Drell--Yan and $J/\psi$ production events are reconstructed 
from the dilepton invariant mass spectrum, where the  $J/\psi$
events correspond to the peak at the invariant mass 
$M = M_{J/\psi} \simeq 3$ GeV, and the genuine Drell--Yan 
events correspond to the continuum spectrum below 
and above the  $J/\psi$ peak.  
For the experimental programs at GSI-HESR, the center of mass
energy is $s = 30-45$ GeV$^2$ in the fixed target mode and $s =
100-200$ GeV$^2$ for the collider option. With  $s = 45$ GeV$^2$
and $s = 200$ GeV$^2$ one has on the $J/\psi$ peak $\tau = x_1 x_2
= M^2/s \simeq 0.2$ and $\tau \simeq 0.05$, respectively. 
In these kinematic
domains valence quarks dominate and the assumptions made above are
justified. 

\begin{figure}[t]
\begin{center}
\scalebox{0.8}{\includegraphics{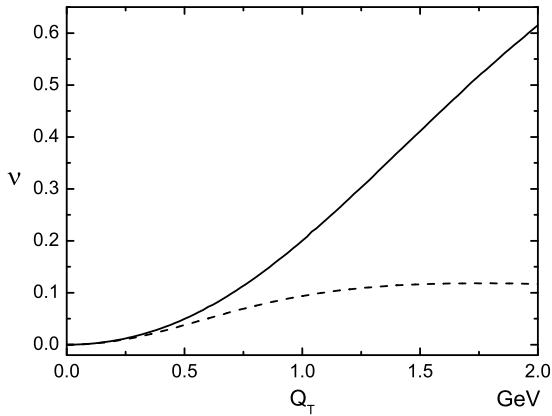}} \caption{\small  The
 $\cos 2 \phi$ azimuthal asymmetry of Drell--Yan production 
in $p\bar{p}$ collisions as a function of 
$Q_T$, for $s = 45$ GeV$^2$ and  
$M^2 = 5$ GeV$^2$. The solid curve is our 
prediction. The dashed curve is the result 
of \cite{bbh03} extended to the kinematic region
of interest here.}
\label{dyfig}
\end{center}
\end{figure}

The $\cos 2 \phi$ asymmetry $\nu$ as a function of $Q_T$ 
is shown in Fig.~\ref{dyfig} for the fixed--target 
GSI kinematics. For a quantitative comparison, in the 
same figure we present the prediction of a model 
with constant quark--spectator coupling \cite{bbh03}, obtained
using the same parameters of \cite{bbh03} but 
for the kinematic domain we are considering here. 
We notice that both the magnitude and the $Q_T$ 
dependence are quite different, similarly to 
what happens in the $\pi N$ Drell--Yan process \cite{lm05}. 
The shape of $\nu$ in the model with a Gaussian 
quark--spectator coupling \cite{GG05} also differs 
from ours and is similar to that of \cite{bbh03}. 

\begin{figure}[t]
\begin{center}
\scalebox{0.8}{\includegraphics{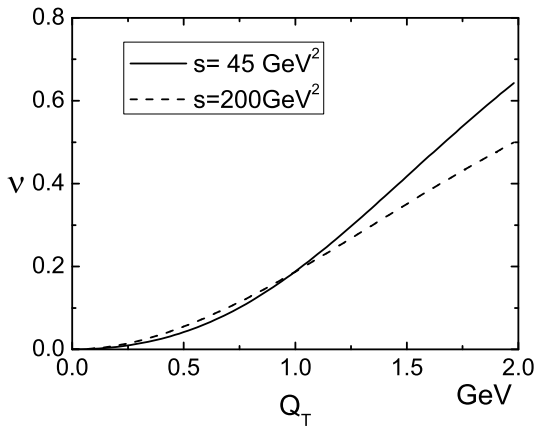}} \caption{\small  The
 $\cos 2 \phi$ azimuthal asymmetry of $p\bar{p}
\rightarrow J/\psi \, X\rightarrow l^+l^- X$ as a function of $Q_T$,
for two values of $s$ in the GSI-HESR kinematic domain.}
\label{jpsifig}
\end{center}
\end{figure}

\begin{figure}[t]
\begin{center}
\scalebox{0.75}{\includegraphics{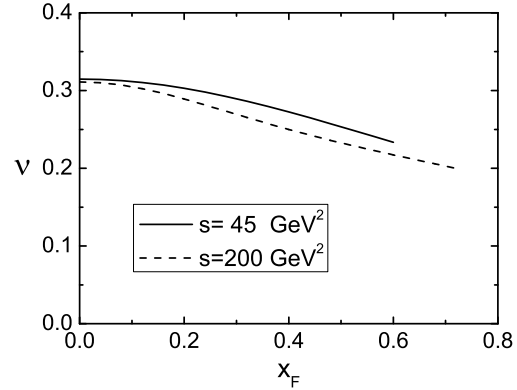}} \caption{\small
The $\cos 2 \phi$ azimuthal asymmetry of $p\bar{p}
\rightarrow J/\psi X\rightarrow l^+l^- X$ as a function of $x_F$,
for two values of $s$ in the GSI-HESR kinematic domain.}
\label{jpsi_xF}
\end{center}
\end{figure}

The $Q_T$-dependence of the $\cos 2 \phi$ asymmetry
$\nu$ in $J/\psi$ production is shown in Fig.~\ref{jpsifig}.
 As one can see, the asymmetry is sizable and increases
with $Q_T$. Some difference between the two curves (corresponding
to two values of $s$) emerges at large $Q_T$, where the asymmetry
is suppressed for the larger center of mass energy.
In Fig.~\ref{jpsi_xF} we plot $\nu$ as a function of $x_F = x_1 -
x_2$. The asymmetry lies in the range 0.2--0.3.

$J/\psi$ production has been recently suggested by Ansel\-mino et
al.~\cite{abdn} as a candidate reaction to probe the nucleon
transversity by measuring the double transverse spin asymmetry in
polarized $p^{\uparrow} \bar{p}^{\uparrow}$ process. The dilepton
production rate on the $J/\psi$ peak is two orders of magnitude
larger than in the continuum region above the $J/\psi$ mass. This
makes $J/\psi$ at GSI energies an excellent process to measure
transverse polarization asymmetries. Here we applied the same idea
to the unpolarized case, and in particular to the $\cos 2 \phi$
asymmetry.
Given the large value of $\nu$ that we found, we can say that the
experimental study of $J/\psi$ and Drell-Yan production from unpolarized
proton-antiproton collisions at GSI represents a very important
source of information on the Boer--Mulders distribution 
(of course, a complete analysis of these processes 
will also require a careful 
consideration of the perturbative effects).

\section{Conclusion}

In summary, we calculated the $\cos 2\phi$ azimuthal asymmetry of
the unpolarized $p\bar{p}$ dilepton production in the continuum 
region and on the $J/\psi$
peak for the GSI kinematics, relying on a model of the
Boer-Mulders function previously adjusted to fit the available
experimental data on $\pi N$ Drell-Yan.
The asymmetry turns out to be rather large, of order 0.2--0.3.
The size of the asymmetry and the high counting rate 
in the $J/\psi$ resonance region 
 make the dilepton production 
in $p \bar p$ scattering at moderate energies
a very promising process to detect the Boer-Mulders function in the early
unpolarized stages of the future GSI experiments.

\begin{acknowledgement}
This work is partially supported by the National
Natural Science Foundation of China (Nos.~10421503, 10575003,
10528510, 10505001), by the Key Grant Project of Chinese Ministry of
Education (No.~305001), by the Research Fund for the Doctoral
Program of Higher Education (China), by the Italian Ministry of
Education, University and Research (PRIN 2003), and by Fondecyt
(Chile) under Project No.~3050047.
\end{acknowledgement}


\begin{thebibliography}{99}


\bibitem{bm}
D.~Boer, P.J.~Mulders, Phys. Rev. {\bf D57} (1998) 5780.

\bibitem{boer}
D.~Boer, Phys. Rev. {\bf D60} (1999) 014012.


\bibitem{bdr}
For a review on transverse polarization phenomena, see V.~Barone,
A.~Drago, P.G.~Ratcliffe, Phys. Rep. {\bf 359} (2002) 1.



\bibitem{na10} NA10 Collaboration, S.~Falciano, et al.,
Z. Phys. {\bf C31} (1986) 513;\\
NA10 Collaboration, M.~Guanziroli, et al., Z. Phys. {\bf C37}
(1988) 545.

\bibitem{conway} E615 Collaboration, J.S.~Conway, et al.,
Phys. Rev. {\bf D39} (1989) 92.


\bibitem{collins02} J.C. Collins,  Phys. Lett.  {\bf B536} (2002) 43.

\bibitem{bjy02}
X.~Ji and F.~Yuan, Phys. Lett. {\bf B543}
(2002) 66;\\
A.V.~Belitsky, X.~Ji, F.~Yuan, Nucl. Phys. {\bf B656} (2003) 165.


\bibitem{gg02}
G.R.~Goldstein and L.~Gamberg, hep-ph/0209085.

\bibitem{bbh03}
D.~Boer, S.J.~Brodsky, D.S.~Hwang, Phys. Rev.  {\bf D67 } (2003)
054003.


\bibitem{lm04} Z.~Lu, B.-Q.~Ma, Phys. Rev.  {\bf D70} (2004)
094044;

\bibitem{lm05} Z.~Lu, B.-Q.~Ma,
Phys. Lett. {\bf B615} (2005) 200.

\bibitem{bsy04}
A.~Bacchetta, A.~Sch\"{a}fer, J.-J.~Yang, Phys. Lett. {\bf B578}
(2004) 109.


\bibitem{gg03} L.P.~Gamberg, G.R.~Goldstein,
K.A.~Oganessyan, Phys. Rev. {\bf D67} (2003) 071504;\\
L.P.~Gamberg, hep-ph/0412367.


\bibitem{BLM05}
V.~Barone, Z.~Lu, B.-Q.~Ma, Phys. Lett. {\bf B632} (2006) 277.


\bibitem{PAX}
PAX Collaboration, V.~Barone, et al., hep-ex/0505054.

\bibitem{PANDA}
PANDA Collaboration, M.~Kotulla, et al., Letter of Intent, 2004.

\bibitem{MC-DY}
A.~Bianconi, M.~Radici, Phys. Rev. {\bf D71} (2005) 074014;\\
A.N.~Sissakian, O.Yu.~Shevchenko, A.P.~Nagaytsev, O.N.~Ivanov,
Phys. Rev. {\bf D72} (2005) 054027.


\bibitem{GG05}
L.P.~Gamberg, G.R.~Goldstein, hep-ph/0506127.



\bibitem{zeus} ZEUS Collaboration, J. Breitweg, et al.,
Phys. Lett. {\bf B481} (2000) 199.

\bibitem{EMC}
EMC Collaboration, M.~Arneodo, et al.,
\Journal{\ZPC}{34}{277}{1987}.


\bibitem{Ji:2004xq}
X.~Ji, J.P.~Ma, F.~Yuan, Phys. Lett. {\bf B597} (2004) 299.

\bibitem{cs77}
J.C.~Collins, D.E.~Soper, Phys. Rev. {\bf D16} (1977) 2219.

\bibitem{collins79}
J.C.~Collins, Phys. Rev. Lett. {\bf 42} (1979) 291.

\bibitem{chiappetta86} 
P.~Chiappetta, M. Le Bellac, Z. Phys. {\bf C32} (1986) 521.

\bibitem{brandenburg93}
A.~Brandenburg, O.~Nachtmann, E.~Mirkes, Z. Phys. 
{\bf C60} (1993) 697. 

\bibitem{barone07}
V.~Barone, B.-Q.~Ma, A.~Prokudin, paper in preparation. 

\bibitem{abdn}
M.~Anselmino, V.~Barone, A.~Drago, N.N.~Nikolaev, Phys. Lett. {\bf
B594} (2004) 97. See also A.V.~Efremov, K.~Goeke, P.~Schweitzer,
Eur. Phys. J. {\bf C35} (2004) 207.


\bibitem{jpsi}
C.E.~Carlson, R.~Suaya, Phys. Rev. {\bf D18} (1978) 760.


\bibitem{sps}
  M.J. Corden,  et al., Phys. Lett. {\bf 68B} (1977) 96;
Phys. Lett.  {\bf 96B} (1980) 411; Phys. Lett. {\bf 98B} (1981)
220.



\bibitem{jmr97} R.~Jakob, P.J.~Mulders, J.~Rodrigues,
Nucl. Phys. {\bf A626} (1997) 937.

   




\end{thebibliography}
\end{document}